\documentclass[iop,apjl,revtex4]{emulateapj}
\usepackage{natbib}
\usepackage{times}
\usepackage{comment}
\newcommand{\eos}{EoS}
\newcommand{\msun}{\ensuremath{M_\sun}}
\newcommand{\msuns}{\ensuremath{M_\sun\,{\rm s}^{-1}}}
\newcommand{\tpb}{\ensuremath{t_{\rm pb}}}
\newcommand{\tadv}{\ensuremath{\tau_{\rm adv}}}
\newcommand{\theat}{\ensuremath{\tau_{\rm heat}}}
\newcommand{\tconv}{\ensuremath{\tau_{\rm conv}}}

\newcommand{\nue}{\ensuremath{\nu_{e}}}
\newcommand{\nuebar}{\ensuremath{\bar \nu_e}}
\newcommand{\numt}{\ensuremath{\nu_{\mu\tau}}}
\newcommand{\numtbar}{\ensuremath{\bar \nu_{\mu\tau}}}
\newcommand{\numu}{\ensuremath{\nu_{\mu}}}
\newcommand{\nutau}{\ensuremath{\nu_{\tau}}}
\newcommand{\numubar}{\ensuremath{\bar \nu_{\mu}}}
\newcommand{\nutaubar}{\ensuremath{\bar \nu_{\tau}}}
\newcommand{\mev}{\mbox{MeV}}

\newcommand{\tgain}{\ensuremath{t_{\rm gain}}}
\newcommand{\dshock}{\ensuremath{d_{\rm shock}}}
\newcommand{\rshock}{\ensuremath{r_{\rm shock}}}
\newcommand{\Ediag}{\ensuremath{E^{+}}}
\newcommand{\Eth}{\ensuremath{E_{\rm th}}}

\newcommand{\Ediagov}{\ensuremath{E^{+}_{\rm ov}}}
\newcommand{\Ediagovrec}{\ensuremath{E^{+}_{\rm ov, rec}}}

\newcommand{\meanE}[1]{\ensuremath{\langle E_{#1}\rangle_{\rm rms}}}
\newcommand{\nusphere}{\ensuremath{\nu\rm-sph}}

\newcommand{\isotope}[2]{$^{#2}$#1}
\newcommand{\gcc}{\ensuremath{{\mbox{g~cm}}^{-3}}}

\newcommand{\kmps}{\ensuremath{\mbox{km~s}^{-1}}}
\newcommand{\cmps}{\ensuremath{\mbox{cm~s}^{-1}}}
\newcommand{\Bethes}{\ensuremath{{\mbox{B~s}}^{-1}}}

\newcommand{\chimera}{{\sc Chimera}}

\slugcomment{}

\shorttitle{Axisymmetric Core-Collapse Supernova Simulations}
\shortauthors{Bruenn et al.}

\begin{document}

\title{Axisymmetric \emph{Ab Initio} Core-Collapse Supernova Simulations of 12--25 \msun\ Stars}

\author{
Stephen W. Bruenn\altaffilmark{1},
Anthony Mezzacappa\altaffilmark{2,3,4},
W. Raphael Hix\altaffilmark{2,3},
Eric J. Lentz\altaffilmark{3,2,5},
O. E. Bronson Messer\altaffilmark{6,3,4},\\
Eric J. Lingerfelt\altaffilmark{2,4},
John M. Blondin\altaffilmark{7},
Eirik Endeve\altaffilmark{4},
Pedro Marronetti\altaffilmark{1,8},
and Konstantin N. Yakunin\altaffilmark{1}
}
\email{bruenn@fau.edu}

\altaffiltext{1}{Department of Physics, Florida Atlantic University, 777 Glades Road, Boca Raton, FL 33431-0991, USA}
\altaffiltext{2}{Physics Division, Oak Ridge National Laboratory, P.O. Box 2008, Oak Ridge, TN 37831-6354, USA}
\altaffiltext{3}{Department of Physics and Astronomy, University of Tennessee, Knoxville, TN 37996-1200, USA}
\altaffiltext{4}{Computer Science and Mathematics Division, Oak Ridge National Laboratory, P.O.Box 2008, Oak Ridge, TN 37831-6164, USA}
\altaffiltext{5}{Joint Institute for Heavy Ion Research, Oak Ridge National Laboratory, P.O. Box 2008, Oak Ridge, TN 37831-6374, USA}
\altaffiltext{6}{National Center for Computational Sciences, Oak Ridge National Laboratory, P.O. Box 2008, Oak Ridge, TN 37831-6164, USA}
\altaffiltext{7}{Department of Physics, North Carolina State University,  Raleigh, NC 27695-8202, USA}
\altaffiltext{8}{Physics Division, National Science Foundation, Arlington, VA 22207 USA}

\begin{abstract}

We present an overview of four \emph{ab initio} axisymmetric core-collapse
supernova simulations employing {\it detailed} spectral neutrino transport
computed with our \chimera\ code and initiated from \citet{WoHe07}
progenitors of mass 12, 15, 20, and 25~\msun.  All four models exhibit
shock revival over $\sim$ 200 ms (leading to the possibility of explosion), 
driven by neutrino  energy deposition.  Hydrodynamic instabilities that impart 
substantial asymmetries to the shock aid these revivals, with convection appearing 
first in the 12~\msun\ model and the standing accretion shock instability (SASI) 
appearing first in the 25 \msun\ model. Three of the models have developed 
pronounced prolate morphologies (the 20~\msun\ model has remained approximately 
spherical).  By 500~ms after bounce the mean shock radii in all four models exceed 
3,000~km and the diagnostic explosion energies are 0.33, 0.66, 0.65, and 0.70~Bethe 
(B~=~$10^{51}$ ergs) for the 12, 15, 20, and 25~\msun\ models, respectively, and are 
increasing. The three least massive of our models are already sufficiently energetic 
to completely unbind the envelopes of their progenitors (i.e., to explode), as
evidenced by our best estimate of their explosion energies, which first become 
positive at 320, 380, and 440 ms after bounce. By 850~ms the 12~\msun\ diagnostic 
explosion energy has saturated at 0.38~B, and our estimate for the final kinetic 
energy of the ejecta is $\sim$ 0.3 B, which is comparable to observations for 
lower-mass progenitors.

\end{abstract}

\keywords{neutrinos --- radiative transfer  --- supernovae: general}

\section{Introduction}

\citet{CoWh66}  originated the idea that core-collapse supernovae (CCSNe) may be neutrino driven. 
Because realistically modeling CCSNe is a computationally complex task of tying together hydrodynamics, neutrino transport, general relativity (GR), nuclear burning, and a realistic equation of state (\eos), and since the many feedbacks and the nonlinearity of the process preclude an analytical treatment, progress in the field has been closely tied to advances in computational resources and code development  \citep{Mezz05, KoSaTa06, Jank12}.
Two decades ago advances of this sort made possible the first generation of CCSNe simulations in two dimensions (2D) \citep{HeBeHi94, BuHaFr95, JaMu96, MeCaBr98b, FrWa02}. 
These simulations, by allowing, among other things, neutrino-driven convection below the shock and continued accretion and shock expansion to coexist, gave explosions. In contrast, continually refined spherically-symmetric CCSNe simulations failed to explode except for the lightest progenitors \citep{RaJa00, MeLiMe01, LiMeTh01, LiMeMe04, BuRaJa03}.
Further investigation has shown the crucial importance of multi-D effects, including the Standing Accretion Shock Instability (SASI) \citep{BlMeDe03}, in shaping explosions, and in accounting for the many observed CCSNe explosion asymmetries \cite[see, e.g.,][]{Jank12}.

While the first-generation 2D CCSNe simulations were limited by their use of gray neutrino transport or precomputed spectral neutrino transport, current computing resources have permitted  second-generation 2D CCSNe simulations with self-consistent, spectral neutrino transport \citep{BuRaJa03, BuRaJa06, BuJaRa06, BuLiDe06, BuLiDe07, BrDiMe06, BrMeHi09b, MaJa09,  SuKoTa10, TaKoSu12, MuJaMa12, MuJaHe12}, essential for accurately computing the neutrino--matter coupling at the heart of the neutrino-driven CCSN mechanism.
Realistic multi-D \emph{ab initio} (neutrino luminosities and spectra computed from {\it detailed} spectral neutrino transport rather than prescribed, parameterized, or incomplete spectral transport) CCSNe simulations are essential for elucidating the details of the neutrino-driven supernova mechanism and fully accounting for supernova observables, including explosion energies, neutrino and gravitational wave signatures, nucleosynthesis, compositional mixing, and neutron star recoils. 
The Garching group has published such simulations in axisymmetry using the Vertex code, incorporating a complete set of neutrino opacities, velocity dependent transport terms, and neutrino-energy-bin coupling in the neutrino transport,  standard \eos s, and GR corrections to Newtonian self-gravity \citep{ BuRaJa06, BuJaRa06, MaJa09} or the conformal flatness GR approximation \citep{MuJaMa12, MuJaHe12} \citep[see][for an overview]{JaHaHu12, MuJaMa12b}. Their successful models either produce weak explosions or were reported before their explosion energies saturated.

\begin{deluxetable*}{lccccc}
\tabletypesize{\scriptsize}
\tablecaption{Model Summary Table\label{tab:models}}
\tablecolumns{6}
\tablewidth{0pt}
\tablehead{
\colhead{} & \multicolumn{5}{c}{Models} \\
\cline{2-6} \\
\colhead{Property} & \multicolumn{2}{c}{B12-WH07} & \colhead{B15-WH07} & \colhead{B20-WH07} & \colhead{B25-WH07} 
}
\startdata
Progenitor mass, $M_{\rm ZAMS}$ [\msun] &\multicolumn{2}{c}{12} & 15 & 20 & 25 \\
Progenitor compactness, $\xi_{1.75}$ &\multicolumn{2}{c}{0.234} & 0.598 & 1.10 & 1.25 \\
Initial simulation domain radius [km] & \multicolumn{2}{c}{30000} & 20000 & 20000 & 20000 \\
Simulation enclosed mass [\msun] &\multicolumn{2}{c}{2.08} & 2.75 & 3.51& 4.57 \\
Radial zone count & \multicolumn{2}{c}{512} & 512 & 512 & 512 \\
Angular zone count & \multicolumn{2}{c}{256} & 256 & 256 & 256 \\
Collapse time (initiation to bounce) [ms] & \multicolumn{2}{c}{266} & 335 & 462 & 470 \\
Bounce, central density [$10^{14}$ \gcc] & \multicolumn{2}{c}{3.13} & 3.45 & 3.42 & 3.29 \\
Bounce, central $Y_e$ &\multicolumn{2}{c}{0.240} & 0.249 & 0.248 & 0.246 \\ 
Bounce, shock position [\msun] &\multicolumn{2}{c}{0.45} & 0.46 & 0.47 & 0.47 \\
Time to growth in gain region mass, $t_{\rm gain}$ [ms] & \multicolumn{2}{c}{256} & 221 & 195 & 192 \\
Time to positive diagnostic energy $>$ 0.01 B [ms] &\multicolumn{2}{c}{219} & 207 & 192 & 197 \\
Time to mean shock radius of 500 km, $t_{500}$ [ms] & \multicolumn{2}{c}{236} & 233 & 208 & 212 \\
Post-bounce epoch, $\tpb$ [ms] & 850 & 500 & 500 & 500 & 500 \\
Post-bounce diagnostic energy, \Ediag\ [B] & {0.38} & 0.33 & 0.66 & 0.65 & 0.70 \\
Post-bounce \Ediagov\ [B] (see text) & {0.29} & 0.0.19 & 0.27 & 0.03 & ---  \\
Post-bounce \Ediagovrec\ [B] (see text) & {0.32} & 0.23 & 0.41 & 0.08 & ---  \\
Post-bounce mean shock radius [km] & {7245} & 3429 & 3293 & 3632 & 3783 \\
Post-bounce shock deformation, \dshock &  {0.42} & 0.31 & 0.52 & 0.027 & 0.31 \\
Post-bounce PNS rest mass [\msun] & 1.48 & 1.48 & 1.66 & 1.80 & 1.88
\enddata
\end{deluxetable*}

The limited number of published \emph{ab initio} CCSNe simulations with detailed neutrino transport, the quantitative differences we obtain relative to the Garching group, and the specifics of our findings (explosions more broadly consistent with observations) motivate 
us to present in this Letter an outline of four simulations being performed with our 
\chimera\ code.  These simulations span progenitors of mass 12--25~\msun\  with detail comparable to \citet{MaJa09}.  Here we describe some of the general features already manifest in our simulations and how they support the viability of the neutrino-driven CCSN mechanism. 

\section{Numerical methods and inputs}
\label{sec:numerical}

\chimera\ is a parallel, multi-physics code built specifically for multidimensional simulation of CCSNe.
It is a combination of separate codes for hydrodynamics and gravity; neutrino transport and opacities; and nuclear EoS and reaction network, coupled by a layer that oversees data management, parallelism, I/O, and control.
The hydrodynamics are evolved via a dimensionally-split, Lagrangian-Remap (PPMLR) scheme  \citep{CoWo84} as implemented in VH1 \citep{HaBlLi12}.
Self-gravity is computed by multipole expansion \citep{MuSt95} replacing the Newtonian monopole  with a GR monopole   \citep[][Case~A]{MaDiJa06}.
Neutrino transport is computed in the ``ray-by-ray-plus'' (RbR+) approximation \citep{BuRaJa03}, where an independent, spherically symmetric transport solve is computed for each angular ``ray'' ($\theta$-zone).
Neutrinos are advected laterally (in the $\theta$-direction) with the fluid and contribute to the lateral pressure gradient where $\rho>10^{12}\,\gcc$.
The transport solver is an improved and updated version of the multi-group flux-limited diffusion transport solver of \citet{Brue85}, enhanced for GR \citep{BrDeMe01}, with an additional geometric flux limiter to prevent the over-rapid transition to free streaming of the standard flux-limiter.  All O(v/c) observer corrections have been included.
We solve for all three flavors of (anti-)neutrinos with four coupled species: \nue, \nuebar, $\numt=\{\numu,\nutau\}$, $\numtbar=\{\numubar,\nutaubar\}$, with 20 energy groups each for $\alpha\epsilon =  4$--250~\mev, where $\alpha$ is the lapse function and $\epsilon$  the comoving-frame group-center energy.
The neutrino--matter interactions include emission, absorption, and non-isoenergetic scattering on free nucleons \citep{RePrLa98} with weak magnetism corrections \citep{Horo02}; emission/absorption (electron capture) on nuclei \citep{LaMaSa03}; isoenergetic scattering on nuclei, including ion-ion correlations; non-isoenergetic scattering on electrons and positrons; and pair emission from $e^+e^-$-annihilation \citep{Brue85} and nucleon-nucleon bremsstrahlung \citep{HaRa98}.
Unlike the Garching group, we do not include the effective mass corrections at high density nor the $\nue\nuebar\rightarrow\numt\numtbar$ process.
We utilize the $K = 220$~\mev\ incompressibility version of the \citet{LaSw91} EoS for  $\rho>10^{11}\,\gcc$ and a modified version of the \cite{Coop85} EoS for  $\rho<10^{11}\,\gcc$ where nuclear statistical equilibrium (NSE) applies.
To aid the transition to the XNet \citep{HiTh99a} 14-species $\alpha$-network ($\alpha$, \isotope{C}{12}-\isotope{Zn}{60}) used for the non-NSE regions, we have constructed a 17-species NSE solver to be used in place of the Cooperstein EoS for $Y_e>0.46$.
An extended version of the Cooperstein electron--photon EoS is used throughout.

During evolution the radial zones are gradually and automatically repositioned during the remap step to track changes in the radial structure.
To minimize restrictions on the time step from the Courant limit,  we ``freeze'' the lateral hydrodynamics for the inner 8~zones during collapse, and, after prompt convection fades, we expand the laterally frozen region to the inner 8--10~km.
In the ``frozen'' region we set $v_\theta = 0$ and skip the lateral hydrodynamic sweep. 
The full radial hydrodynamics and neutrino transport are always computed to the center of the simulation for all angular rays. 
A more extensive description of \chimera\ is under preparation.

Four non-rotating axisymmetric models (designated B12-WH07, B15-WH07, B20-WH07, B25-WH07, corresponding to progenitor masses of 12, 15, 20, and 25~\msun), and identical 1D models, are initialized without artificial perturbation from the pre-supernova progenitors of \citet{WoHe07}. 
A grid of 512 non-equally spaced radial zones covers from the stellar center into the oxygen-rich layers. 
The 2D models employ 256 uniform-sized angular zones from 0 to $\pi$.

\begin{figure*}
\epsscale{1.1}
\plotone{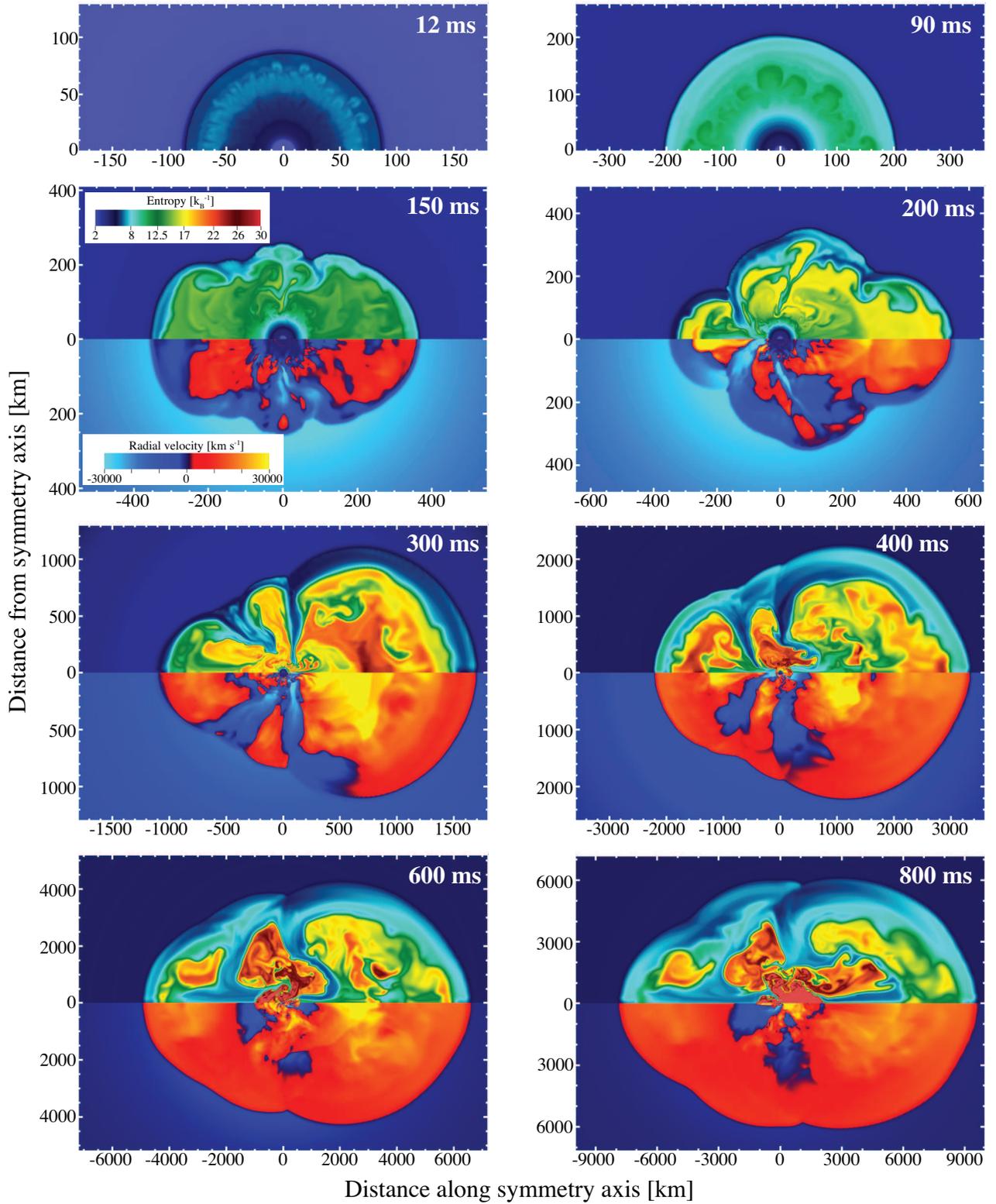}
\caption{Evolution of the entropy (upper half) and radial velocity (lower half)  for B12-WH07, with snapshots at $\tpb=12$, 90, 150, 200, 300, 400, 600,  and 800~ms.  The  scale grows in time to capture the expansion of the supernova shockwave, but the colormaps remain constant. The radial velocity portion is omitted for the first two snapshots. (An animated version of this plot is available at \texttt{http://astro.phys.utk.edu/activities\%3achimera\%3aseriesb})\label{fig:entropy}}.
\end{figure*}

\section{Results}
\label{sec:results}

All 2D models experience similar evolution during collapse and immediately after bounce. 
Theta velocities of a few \cmps\ develop by bounce, most likely from numerical roundoff. 
These grow to at most $\sim10\,\kmps$ and $\sim100\,\kmps$ in the preshock and postshock regions, respectively, at a post-bounce time, \tpb, of 10~ms.
At $\tpb\sim2$~ms the shock breaks through the \nue-sphere, initiating a rapid deleptonization front whose inner edge advects inward with the fluid. 
An unstable lepton gradient results, which is stabilized by the positive entropy gradient laid down by the shock as it initially strengthens. 
The initial oscillations of the proto-neutron star (PNS) periodically strengthen and weaken the shock, causing entropy-unstable pockets and transient convective episodes behind the shock. 
By $\tpb\sim12$~ms an extended unstable entropy gradient has developed behind the shock as it weakens, extending down to $\sim60$~km and driving a brief episode of convection (Figure~\ref{fig:entropy}; 12~ms panel). This episode of ``prompt'' convection has little appreciable effect on the shock, and by $\tpb\sim40$~ms convective activity has ceased.

\begin{figure}
\epsscale{1.1}
\plotone{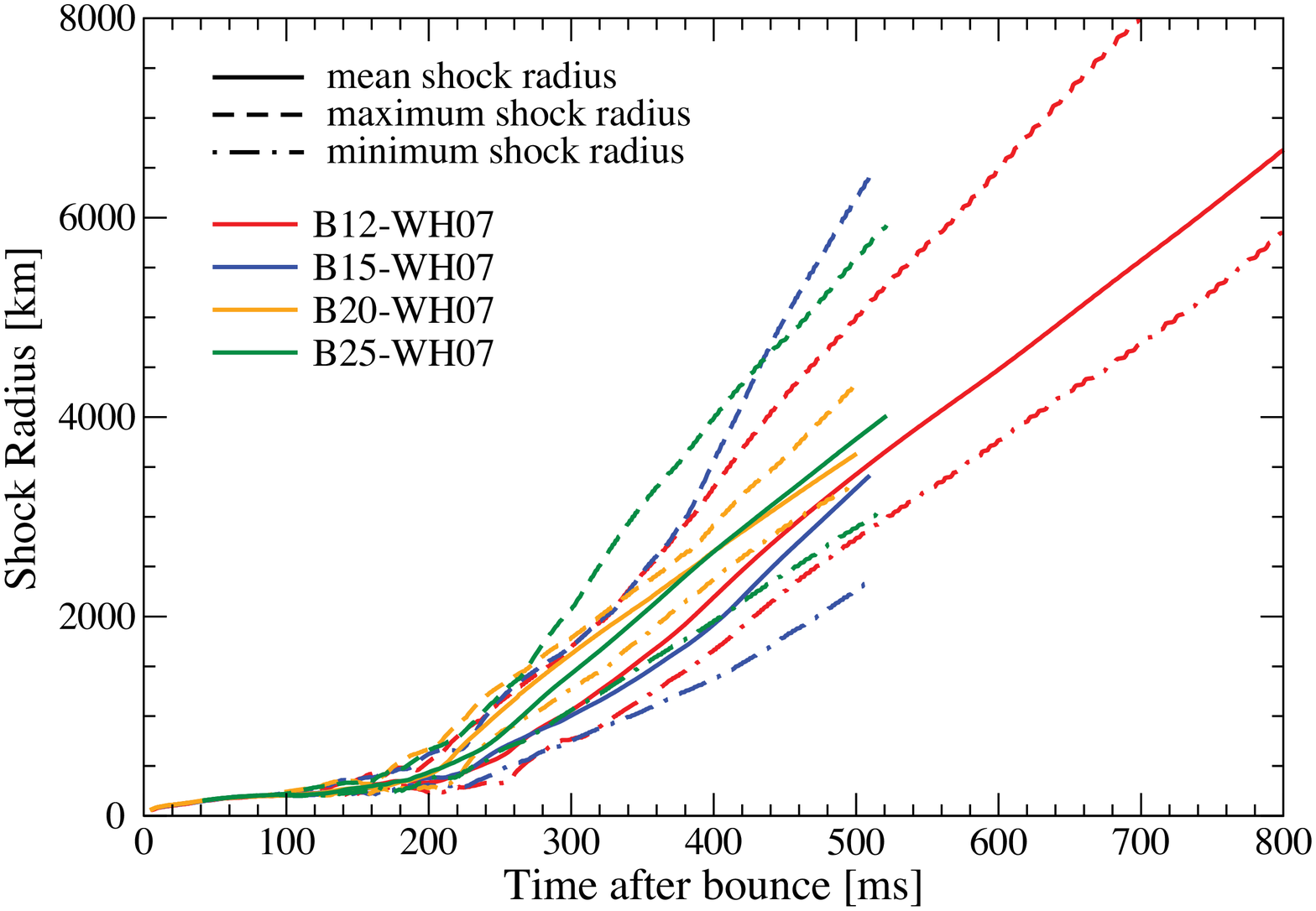}
\plotone{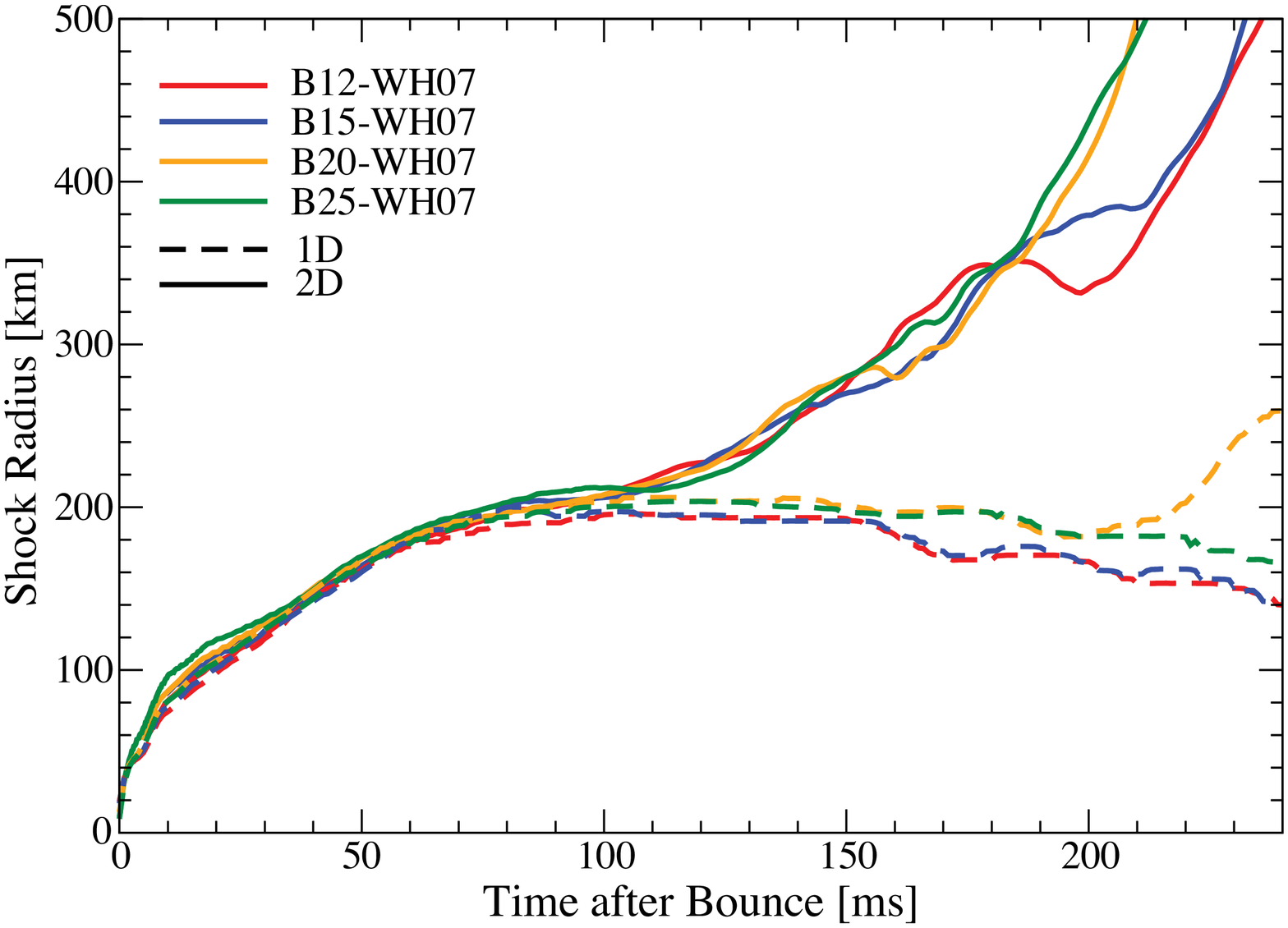}
\caption{Upper panel: Mean, minimum, and maximum shock radii versus \tpb\ for all simulations. Lower panel: Mean shock radii for all simulations and their 1D equivalents. \label{fig:shocks}}
\end{figure}

\begin{figure}
\epsscale{1.2}
\plotone{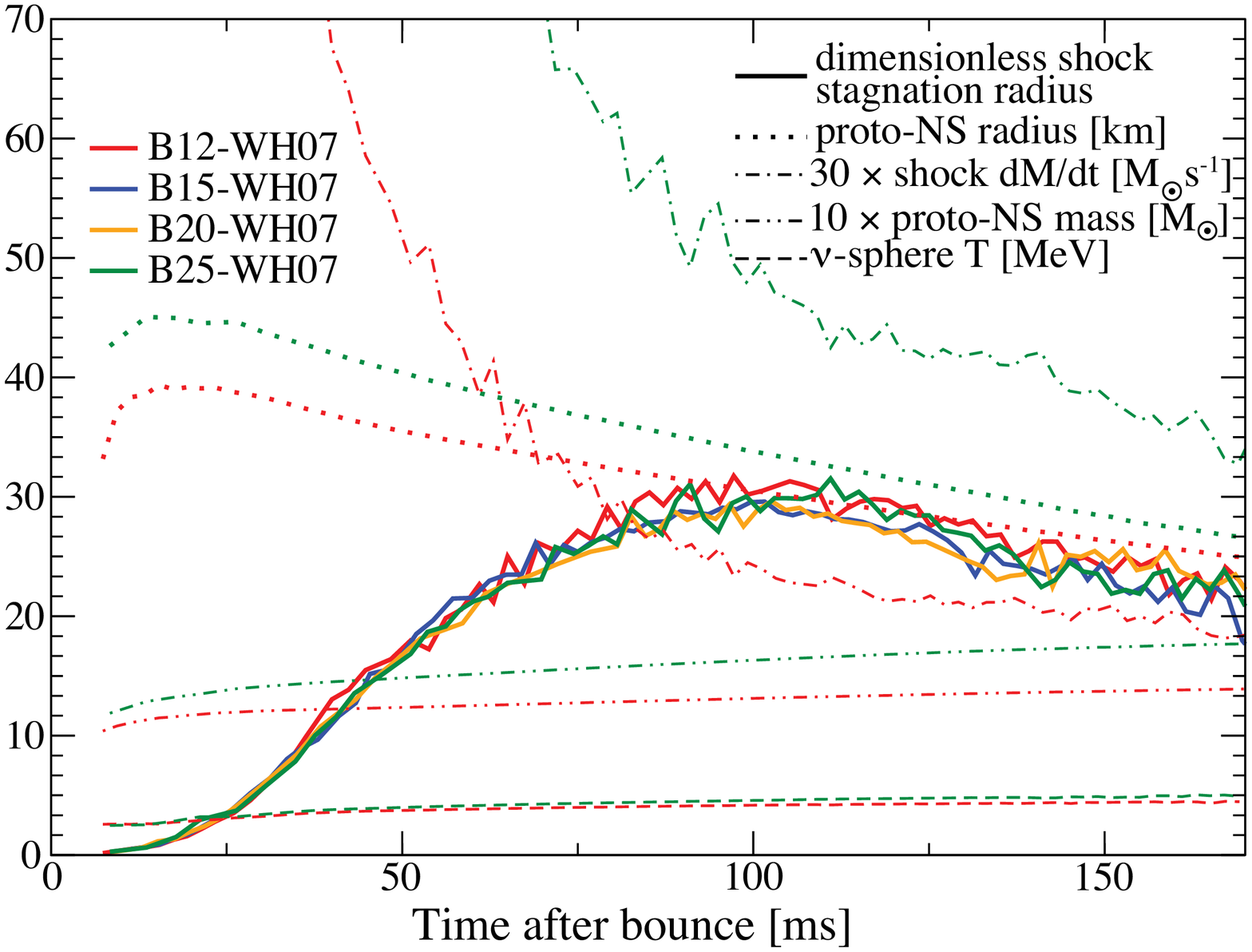}
\plotone{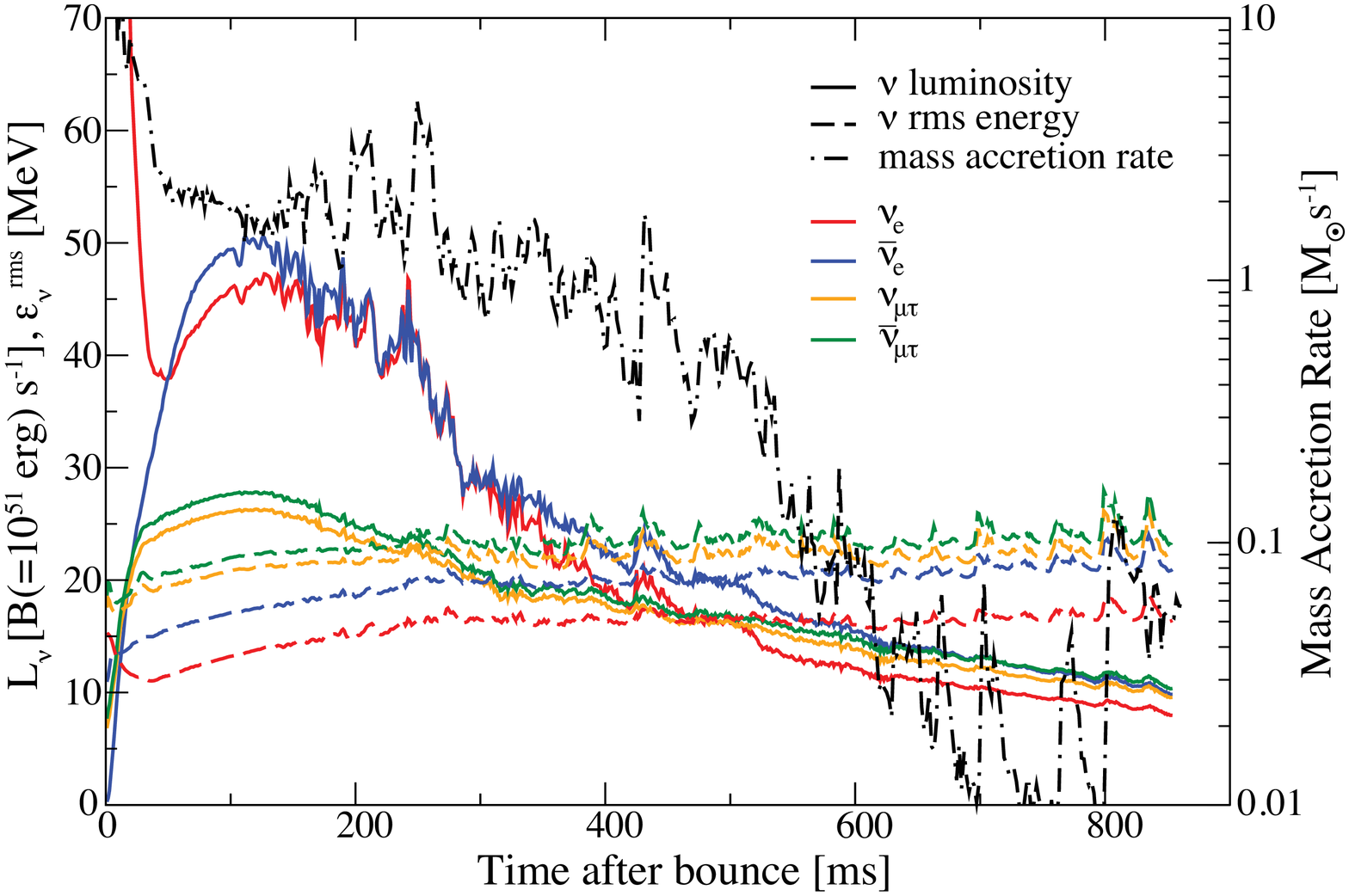}
\caption{Upper panel: Analytic shock using Equation~1 of \citet{Jank12} (solid lines, dimensionless) for all simulations; and PNS radii ($\rho=10^{11}\gcc$ surface; dotted, km), $30\times dM/dt$ at shock (dash-dotted, \msuns), proto-NS mass (double-dot-dashed, 0.1~\msun), and $\nu$-sphere temperature (dashed, \mev) for the extreme cases, B12-WH07 and B25-WH07.   Lower panel: Luminosity (solid lines, \Bethes) and comoving-frame rms energies (dashed, \mev) for all species of neutrinos, and mass accretion rate onto the PNS (dot-dashed, \msuns) for  B12-WH07.\label{fig:analytic}}
\end{figure}

Up to $\tpb\sim100$~ms, the shock trajectories of corresponding 1D and 2D models track each other closely (Figure~\ref{fig:shocks}; lower panel).
Remarkably, the shock trajectories for different progenitors track each other rather closely as well.
This is a consequence of the compensating factors that determine the radius of the shock during its quasi-stationary accretion phase.
Referring to Equation~1 of \cite{Jank12}, these factors (graphed for B12-WH07 and B25-WH07 in Figure~\ref{fig:analytic}; upper panel) are the PNS radius and the $\nu$-sphere temperature, whose increase increases the shock radius, and the PNS mass and the mass accretion rate at the shock whose increase has the opposite effect.
Despite the variations of these factors, their combination during the quasi-stationary accretion phase (Figure~\ref{fig:analytic}; upper panel, solid lines) are quite similar.

At $\tpb\sim100$~ms several evolutionary developments occur in all  2D models, causing the shock trajectories of the 1D and 2D models to diverge. 
First, the ratio of \tadv, the advection timescale through the gain region (region of positive net neutrino heating below the shock), to \tconv, the Brunt-V\"{a}is\"{a}la convective growth timescale, begins to exceed the critical value  $\tadv/\tconv\gtrsim3$ \citep{FoScJa06}. 
Convection in the gain  region  begins (Figure~\ref{fig:entropy}; 90~ms panel), driven by neutrino heating and seeded by the prompt convection episode. 
Neutrino-driven convection pushes the shock out, enlarging the gain region, and increasing \tadv. 
Second, the ratio of \tadv\ to \theat\ (ratio of the thermal energy \Eth\ to the net neutrino heating rate $\dot{Q}_{\nue}$ in the gain region) begins to exceed unity. 
(Using the total energy rather than \Eth\ to compute \theat\ increases by $\sim30$\% the time from bounce for \tadv/\theat\  to exceed unity.)
This signals the beginning of rising entropy in the gain region and an eventual shock revival  \citep{BuJaRa06,ThQuBu05}. 
Third, the SASI begins to manifest through low-mode distortions of the shock and oscillating patterns of the radial velocity (evident in the animated version of Figure~\ref{fig:entropy}), causing, for example, the gain region to swell along the poles and contract in the equatorial regions (Figure~\ref{fig:entropy}; 150~ms panel). 
This distortion mode increases  $\tadv/\theat$ in the expanded regions and decreases it in the contracted regions creating outflow plumes and inflow funnels in the former and latter regions, respectively. 
Thus, while the shocks of the 1D models  slowly contract after $\tpb\sim100$~ms, the average shock radii of the 2D models continue to slowly increase. The expansion accelerates until $\tpb\sim200$~ms (Figure~\ref{fig:shocks}).

Going from B12-WH07 to B25-WH07, $\dot{Q}_{\nue}$ increases  $\sim2$--$2.5$ fold, $E_{\rm th}$ increases  $\sim1.5$ fold, and \tadv\ decreases  $\sim1.5$ fold, consequently  $\tadv/\theat\gtrsim1$ at about the same \tpb\ for all four models. 
On the other hand, the decrease in \tadv\  noticeably delays the onset of neutrino-driven convection in the more massive models -- from $\tpb\sim60$~ms (B12-WH07) to  $\tpb\sim100$~ms (B25-WH07). 
Consequently, it appears that neutrino-driven convection precedes the SASI in B12-WH07 but  follows it in the B25-WH07, the SASI having time to saturate in the latter model before the onset of convection \citep[cf.][]{MuJaHe12}. B15-WH07 and B20-WH07 are intermediate cases and precedence of convection or the SASI is not clear.

A Legendre decomposition of the shock deformation \citep[computed per][]{BlMe06} indicates that the growth of the SASI is dominated by the $l = 2$ (quadrapolar) mode in B12-WH07 and B15-WH07, while the $l=1$ (dipolar) mode is dominant in B20-WH07 and B25-WH07. The subsequent rise in the amplitude of shock deformation as the shock begins to accelerate outward is always dominated by the $l=1$ mode.
The shock deformation oscillates, with a period increasing with the progenitor mass, from 18~ms for B12-WH07 to 30~ms for B25-WH07, though period variations are seen in each model.

The neutrino luminosities $L_{\nu}$ and rms energies \meanE{\nu} for all models follow a similar pattern to that of B12-WH07 (Figure~\ref{fig:analytic}; lower panel).
Following the \nue-break-out burst, the luminosities of all neutrino species peak between 100 and 200~ms.
The \nue-, \nuebar-luminosities, which arise both from the core and from the energy released by accreting matter, exhibit a more pronounced peak during the peak of the mass accretion rate than $L_{\numt}$ and $L_{\numtbar}$, which arise more exclusively from the core. After 200~ms there is a rapid falloff in $L_{\nue}$ and $L_{\nuebar}$ as the shock begins to accelerate outward and the mass accretion rate declines. 
\meanE{\nu} follows the usual hierarchy, with energy increasing from \nue\ to \nuebar\ to \numt\ to \numtbar, the latter three becoming separated by only a few MeV after several hundred ms. The split between  \meanE{\numt}\ and \meanE{\numtbar}\ is due to weak magnetism, which increases (decreases) the opacities of  \numt\ (\numtbar).
Weak magnetism also causes the \nuebar-luminosity to exceed the \nue-luminosity at times after after bounce for this model, though the \nue\ number luminosity is always dominant..
\meanE{\nu} for all neutrino species increases slowly as the $\nu$-spheres contract and heat along with the PNS.

Progenitor masses are correlated with the compactness parameter of \cite{OcOt11} (see Table~\ref{tab:models}), and the luminosities of all neutrino flavors increase with this parameter, nearly doubling from B12-WH07 to B25-WH07 between 50 and 250~ms after bounce. 
\meanE{\nu}, however, shows only a small increase (a few MeV) with compactness. 
These trends can be attributed to the dependence of the luminosity on $R^{2}_{\nusphere}T^{4}_{\nusphere}$ and the dependence of the \meanE{\nu} only on $T_{\nusphere}$, where $R_{\nusphere}$ and $T_{\nusphere}$ are radius and temperature of the $\nu$-sphere, both of which increase with the progenitor compactness during the accretion shock phase (Figure~\ref{fig:analytic}, upper panel;  note  $R_{\nusphere}\propto R_{\rm PNS}$). 

For all the models, $L_{\nu}$ and \meanE{\nu} are initially isotropic.
By $\tpb=60$~ms and $\tpb=100$~ms for the lower and higher mass models, respectively, anisotropies $\sim$10\% in $L_{\nue,\nuebar}$ appear on an $l \sim10$ scale, the scale of the 
entropy-driven convection appearing for these models at that time. The $L_{\nue}$ ($L_{\nuebar}$) maxima are anti-correlated (correlated) with the convective plumes, the plumes having higher entropy but lower electron fraction (higher positron density).
By $\tpb=120$~ms anisotropies exist at the $\sim$~30\% level in all models on a scale of $l=2$--3, with $L_{\nue}$ and $L_{\nuebar}$ maxima becoming more correlated with each other and with the convective plumes and accretion funnels that frequently curve beneath them.
By $\tpb =200$~ms, and thereafter, there is a strong $l=1$ and $l=2$ component in the $L_{\nue,\nuebar}$ anisotropy, again correlated with the morphology of the expanding hot plumes and accretion funnels (Figure~\ref{fig:entropy}; bottom four panels).
$L_{\numt,\numtbar}$ remain quite isotropic up to 200~ms after bounce, but follow, with smaller amplitude, the anisotropic pattern of $L_{\nue,\nuebar}$ thereafter.

At $\tpb\sim200$~ms several parameters indicate imminent shock revival and the onset of  explosion for each model. 
First, around this time (noted as \tgain\ in Table \ref{tab:models}), the mass residing in the gain region begins to rapidly increase, which increases \tadv\  and heating efficiency \citep{Jank01, MuBu08, MaJa09, HaMaMu12}. 
Second, the  $\tadv/\theat$ rises significantly above unity around this time, signaling shock revival \citep{BuJaRa06,ThQuBu05}. 
Third, near this time, the mean shock radius exceeds 500 km (denoted $t_{500}$ in Table \ref{tab:models}).

A useful parameter characterizing the morphology of the expanding shock is the shock deformation parameter,\begin{equation} 
\dshock= \frac{\max(\rshock(\theta)\cos\theta)-\min(\rshock(\theta)\cos\theta)}{2\times \max(\rshock(\theta)\sin\theta)}-1
\label{eq:1}
\end{equation}
\citep{ScKiJa06}, where $\rshock(\theta)$ is the shock radius as a function of the polar angle, $\theta$.
Prolate, oblate, and spherical shocks have positive, negative, and vanishing values of \dshock, respectively.
Values of \dshock\ for the models at $\tpb=500$~ms (Table~\ref{tab:models}) indicate that the shock in three models is significantly prolate while for B20-WH07 it is nearly spherical (see also Figure \ref{fig:shocks}, upper panel). 

\begin{figure}
\epsscale{1.0}
\plotone{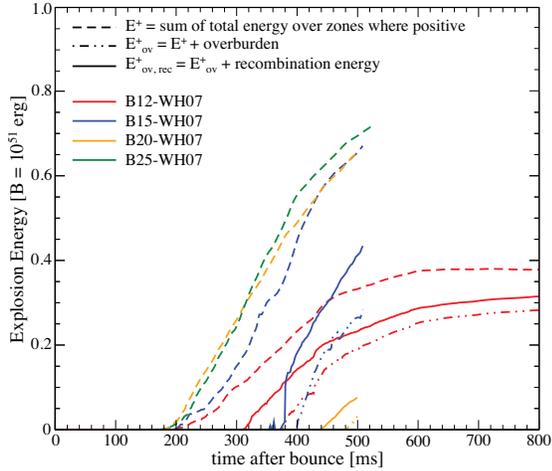}
\caption{Diagnostic energy (\Ediag; dashed lines) versus \tpb\ for all simulations. Dash-dotted lines (\Ediagov) include binding energy of overburden and dashed lines (\Ediagovrec) also include estimated energy gain from nuclear recombination to \isotope{Ni}{56} (see definitions in text).\label{fig:energy}}
\end{figure}

A measure of the explosion energy frequently used is the diagnostic energy, \Ediag\ \citep[integral of total energy, = kinetic + internal + gravitational, where positive;][]{BuJaRa06,SuKoTa10, MuJaMa12}. 
In Figure~\ref{fig:energy} we show \Ediag\ for our models, as well as two other measures of the explosion energy: $\Ediagov=\Ediag+$ overburden, and $\Ediagovrec=\Ediagov+$ recombination energy. 
The overburden is the total energy of all negative energy zones on the grid above the innermost positive energy zones plus the total energy of the off-grid material; the recombination energy is the energy that would be released if all \isotope{He}{4} and neutron--proton pairs recombined to \isotope{Ni}{56} in the positive total energy zones.  We expect that \Ediagov\ and \Ediagovrec\ bound the eventual observable explosion energy from below and above, respectively.

At 500~ms all three explosion energy measures are positive, and increasing, for B12-WH07, B15-WH07, and B20-WH07, but only \Ediag\ is positive for B25-WH07 due to its large off-grid binding energy of -0.655~B. 
For B12-WH07 at 850~ms, the post-shock velocities exceed escape velocity over one quarter of the shock surface; the accretion rate onto the PNS is essentially zero; and the model is beginning to transit from the simultaneous accretion and outflow phase to the wind phase. 
The B12-WH07 explosion energies at 850~ms are $\Ediag=0.38$~B, $\Ediagov=0.285$~B, and $\Ediagovrec=0.32$~B and appear to be leveling off.  These energies are broadly consistent with observations for lower mass progenitors \citep{Smar09}. 

\section{Discussion}
\label{sec:discussion}

We have presented results of \emph{ab initio} axisymmetric core-collapse supernova simulations with \emph{detailed} spectral neutrino transport for a suite of four nonrotating models spanning the mass range 12--25~\msun, through $\tpb=500$~ms for B15-WH07, B20-WH07, and B25-WH07, and $\tpb=850$~ms for B12-WH07. 
At 500~ms, all models show clear indications of developing neutrino-driven explosions, aided initially by strong convective and SASI activity, which imparts pronounced prolate shock deformations in B12-WH07, B15-WH07, and B25-WH07, indicating that large explosion asymmetries are probable (in agreement with observation).

The development of explosions in these models, taking 208--236~ms for the mean shock to reach 500~km, is much slower than in older models---for example, the delayed model of Wilson used by \citet{WoWiMa94}, for which $t_{500}\sim150$~ms, or the 2D gray neutrino transport models of \citet{HeBeHi94}, for which $t_{500}\sim75$~ms. 
The explosion development times of the \chimera\ models reported here are comparable to those of the Garching group for their models that use modern progenitors ($t_{500}\sim250$~ms for model G11 from \citet{MuJaMa12} and $t_{500}\sim230$~ms for model s27 from \citet{MuJaHe12}), but significantly shorter than for their models that use older progenitors ($t_{500}\sim600$~ms for model G15 of \citet{MuJaMa12}). 
This suggests a significant progenitor dependence, which makes detailed comparisons difficult. 
The most pronounced difference between our models and those of the Garching group is the growth of the explosion energy.  
For example, once the explosion develops, \Ediag\ for both their G15 and s27 models \citep{JaHaHu12} grow at a rate of $\sim 0.5$~\Bethes, a quarter of the rate in our model B15-WH07.  
The diagnostic energy in the G11 model saturates at $\Ediag\sim0.03$, one tenth of the value of our model B12-WH07.  
Future simulations with both groups using identical progenitors will help elucidate the source of these differences.

Final conclusions regarding the viability of the neutrino-driven CCSN mechanism must wait until different groups obtain similar results from detailed \emph{ab initio} simulations initiated from the same progenitors and carried out in 3D. Our simulations thus far support the viability of the neutrino-driven supernova mechanism for the lower progenitor masses and are consistent with observations. 
These simulations are continuing, and we will publish more complete analyses as they complete.

\acknowledgements
We thank H.-Th. Janka and B. M{\"u}ller for a very careful reading of this manuscript and helpful comments.
This research was supported by the U.S. Department of Energy Offices of Nuclear Physics and Advanced Scientific Computing Research;
 the NASA Astrophysics Theory and Fundamental Physics Program (grants NNH08AH71I and NNH11AQ72I); and the National Science Foundation PetaApps Program (grants  OCI-0749242, OCI-0749204, and OCI-0749248).
This research was also supported by
the NSF  through TeraGrid resources provided by the National Institute for Computational Sciences under grant number TG-MCA08X010; 
resources of the National Energy Research Scientific Computing Center, supported by the U.S. DoE Office of Science under Contract No. DE-AC02-05CH11231; and
an award of computer time from the Innovative and Novel Computational Impact on Theory and Experiment (INCITE) program at the Oak Ridge Leadership Computing Facility, supported by the  U.S. DoE Office of Science under Contract No. DE-AC05-00OR22725.

%\bibliographystyle{apj}
%\bibliography{apj_journals,add_journals,supernova,network,hydro-mhd,rspn_process}

\begin{thebibliography}{52}
\expandafter\ifx\csname natexlab\endcsname\relax\def\natexlab#1{#1}\fi

\bibitem[{Blondin {et~al.}(2003)Blondin, Mezzacappa, \& DeMarino}]{BlMeDe03}
Blondin, J., Mezzacappa, A., \& DeMarino, C. 2003, ApJ, 584, 971

\bibitem[{{Blondin} \& {Mezzacappa}(2006)}]{BlMe06}
{Blondin}, J.~M. \& {Mezzacappa}, A. 2006, ApJ, 642, 401

\bibitem[{Bruenn(1985)}]{Brue85}
Bruenn, S.~W. 1985, ApJS, 58, 771

\bibitem[{{Bruenn} {et~al.}(2001){Bruenn}, {De Nisco}, \&
  {Mezzacappa}}]{BrDeMe01}
{Bruenn}, S.~W., {De Nisco}, K.~R., \& {Mezzacappa}, A. 2001, ApJ, 560, 326

\bibitem[{Bruenn {et~al.}(2006)Bruenn, Dirk, Mezzacappa, Hayes, Blondin, Hix,
  \& Messer}]{BrDiMe06}
Bruenn, S.~W., Dirk, C.~J., Mezzacappa, A., Hayes, J.~C., Blondin, J.~M., Hix,
  W.~R., \& Messer, O. E.~B. 2006, J. Phys.: Conf. Ser., 46, 393

\bibitem[{{Bruenn} {et~al.}(2009){Bruenn}, {Mezzacappa}, {Hix}, {Blondin},
  {Marronetti}, {Messer}, {Dirk}, \& {Yoshida}}]{BrMeHi09b}
{Bruenn}, S.~W., {Mezzacappa}, A., {Hix}, W.~R., {Blondin}, J.~M.,
  {Marronetti}, P., {Messer}, O.~E.~B., {Dirk}, C.~J., \& {Yoshida}, S. 2009,
  J. Phys.: Conf. Ser., 180, 012018

\bibitem[{{Buras} {et~al.}(2006{\natexlab{a}}){Buras}, {Janka}, {Rampp}, \&
  {Kifonidis}}]{BuJaRa06}
{Buras}, R., {Janka}, H.-T., {Rampp}, M., \& {Kifonidis}, K.
  2006{\natexlab{a}}, A\&A, 457, 281

\bibitem[{{Buras} {et~al.}(2003){Buras}, {Rampp}, {Janka}, \&
  {Kifonidis}}]{BuRaJa03}
{Buras}, R., {Rampp}, M., {Janka}, H.-T., \& {Kifonidis}, K. 2003, Phys. Rev.
  Lett., 90, 241101

\bibitem[{{Buras} {et~al.}(2006{\natexlab{b}}){Buras}, {Rampp}, {Janka}, \&
  {Kifonidis}}]{BuRaJa06}
{Buras}, R., {Rampp}, M., {Janka}, H.-T., \& {Kifonidis}, K. 2006{\natexlab{b}}, A\&A, 447, 1049

\bibitem[{Burrows {et~al.}(1995)Burrows, Hayes, \& Fryxell}]{BuHaFr95}
Burrows, A., Hayes, J., \& Fryxell, B.~A. 1995, ApJ, 450, 830

\bibitem[{{Burrows} {et~al.}(2006){Burrows}, {Livne}, {Dessart}, {Ott}, \&
  {Murphy}}]{BuLiDe06}
{Burrows}, A., {Livne}, E., {Dessart}, L., {Ott}, C.~D., \& {Murphy}, J. 2006,
  ApJ, 640, 878

\bibitem[{{Burrows} {et~al.}(2007){Burrows}, {Livne}, {Dessart}, {Ott}, \&
  {Murphy}}]{BuLiDe07}
{Burrows}, A., {Livne}, E., {Dessart}, L., {Ott}, C.~D., \& {Murphy}, J. 2007, ApJ, 655, 416

\bibitem[{{Colella} \& {Woodward}(1984)}]{CoWo84}
{Colella}, P. \& {Woodward}, P. 1984, J. Comp. Phys., 54, 174

\bibitem[{{Colgate} \& {White}(1966)}]{CoWh66}
{Colgate}, S.~A. \& {White}, R.~H. 1966, ApJ, 143, 626

\bibitem[{{Cooperstein}(1985)}]{Coop85}
{Cooperstein}, J. 1985, Nucl. Phys. A, 438, 722

\bibitem[{{Foglizzo} {et~al.}(2006){Foglizzo}, {Scheck}, \& {Janka}}]{FoScJa06}
{Foglizzo}, T., {Scheck}, L., \& {Janka}, H.-T. 2006, ApJ, 652, 1436

\bibitem[{{Fryer} \& {Warren}(2002)}]{FrWa02}
{Fryer}, C.~L. \& {Warren}, M.~S. 2002, ApJ, 574, L65

\bibitem[{{Hanke} {et~al.}(2012){Hanke}, {Marek}, {Mueller}, \&
  {Janka}}]{HaMaMu12}
{Hanke}, F., {Marek}, A., {Mueller}, B., \& {Janka}, H.-T. 2012, ApJ, 755, 138

\bibitem[{{Hannestad} \& {Raffelt}(1998)}]{HaRa98}
{Hannestad}, S. \& {Raffelt}, G. 1998, ApJ, 507, 339

\bibitem[{{Hawley} {et~al.}(2012){Hawley}, {Blondin}, {Lindahl}, \&
  {Lufkin}}]{HaBlLi12}
{Hawley}, J., {Blondin}, J., {Lindahl}, G., \& {Lufkin}, E. 2012, Astrophysics
  Source Code Library, 4007

\bibitem[{Herant {et~al.}(1994)Herant, Benz, Hix, Fryer, \& Colgate}]{HeBeHi94}
Herant, M., Benz, W., Hix, W.~R., Fryer, C.~L., \& Colgate, S.~A. 1994, ApJ,
  435, 339

\bibitem[{{Hix} \& {Thielemann}(1999)}]{HiTh99a}
{Hix}, W.~R. \& {Thielemann}, F.-K. 1999, ApJ, 511, 862

\bibitem[{{Horowitz}(2002)}]{Horo02}
{Horowitz}, C.~J. 2002, Phys. Rev. D, 65, 43001

\bibitem[{{Janka}(2001)}]{Jank01}
{Janka}, H.-T. 2001, A\&A, 368, 527

\bibitem[{{Janka}(2012)}]{Jank12}
{Janka}, H.-T. 2012, Annu. Rev. Nucl. Part. Sci., 62, 407

\bibitem[{{Janka} {et~al.}(2012){Janka}, {Hanke}, {Huedepohl}, {Marek},
  {M{\"u}ller}, \& {Obergaulinger}}]{JaHaHu12}
{Janka}, H.-T., {Hanke}, F., {Huedepohl}, L., {Marek}, A., {M{\"u}ller}, B., \&
  {Obergaulinger}, M. 2012, ArXiv e-prints

\bibitem[{Janka \& M{\"u}ller(1996)}]{JaMu96}
Janka, H.-T. \& M{\"u}ller, E. 1996, A\&A, 306, 167

\bibitem[{{Kotake} {et~al.}(2006){Kotake}, {Sato}, \& {Takahashi}}]{KoSaTa06}
{Kotake}, K., {Sato}, K., \& {Takahashi}, K. 2006, Reports on Progress in
  Physics, 69, 971

\bibitem[{{Langanke} {et~al.}(2003){Langanke}, {Mart{\'\i}nez-Pinedo},
  {Sampaio}, {Dean}, {Hix}, {Messer}, {Mezzacappa}, {Liebend{\"o}rfer},
  {Janka}, \& {Rampp}}]{LaMaSa03}
{Langanke}, K., {Mart{\'\i}nez-Pinedo}, G., {Sampaio}, J.~M., {Dean}, D.~J.,
  {Hix}, W.~R., {Messer}, O.~E., {Mezzacappa}, A., {Liebend{\"o}rfer}, M.,
  {Janka}, H.-T., \& {Rampp}, M. 2003, Phys. Rev. Lett., 90, 241102

\bibitem[{Lattimer \& Swesty(1991)}]{LaSw91}
Lattimer, J. \& Swesty, F.~D. 1991, Nucl. Phys. A, 535, 331

\bibitem[{{Liebend{\"o}rfer} {et~al.}(2004){Liebend{\"o}rfer}, {Messer},
  {Mezzacappa}, {Bruenn}, {Cardall}, \& {Thielemann}}]{LiMeMe04}
{Liebend{\"o}rfer}, M., {Messer}, O.~E.~B., {Mezzacappa}, A., {Bruenn}, S.~W.,
  {Cardall}, C.~Y., \& {Thielemann}, F.-K. 2004, ApJS, 150, 263

\bibitem[{{Liebend{\"o}rfer} {et~al.}(2001){Liebend{\"o}rfer}, {Mezzacappa},
  {Thielemann}, {Messer}, {Hix}, \& {Bruenn}}]{LiMeTh01}
{Liebend{\"o}rfer}, M., {Mezzacappa}, A., {Thielemann}, F.-K., {Messer}, O.
  E.~B., {Hix}, W.~R., \& {Bruenn}, S.~W. 2001, Phys. Rev. D, 63, 103004

\bibitem[{{Marek} {et~al.}(2006){Marek}, {Dimmelmeier}, {Janka}, {M{\"u}ller},
  \& {Buras}}]{MaDiJa06}
{Marek}, A., {Dimmelmeier}, H., {Janka}, H.-T., {M{\"u}ller}, E., \& {Buras},
  R. 2006, A\&A, 445, 273

\bibitem[{{Marek} \& {Janka}(2009)}]{MaJa09}
{Marek}, A. \& {Janka}, H.-T. 2009, ApJ, 694, 664

\bibitem[{{Mezzacappa}(2005)}]{Mezz05}
{Mezzacappa}, A. 2005, Annu. Rev. Nucl. Part. Sci., 55, 467

\bibitem[{Mezzacappa {et~al.}(1998)Mezzacappa, Calder, Bruenn, Blondin, Guidry,
  Strayer, \& Umar}]{MeCaBr98b}
Mezzacappa, A., Calder, A.~C., Bruenn, S.~W., Blondin, J.~M., Guidry, M.~W.,
  Strayer, M.~R., \& Umar, A.~S. 1998, ApJ, 495, 911

\bibitem[{{Mezzacappa} {et~al.}(2001){Mezzacappa}, {Liebend{\"o}rfer},
  {Messer}, {Hix}, {Thielemann}, \& {Bruenn}}]{MeLiMe01}
{Mezzacappa}, A., {Liebend{\"o}rfer}, M., {Messer}, O.~E., {Hix}, W.~R.,
  {Thielemann}, F.-K., \& {Bruenn}, S.~W. 2001, Phys. Rev. Lett., 86, 1935

\bibitem[{{M{\"u}ller} {et~al.}(2012{\natexlab{a}}){M{\"u}ller}, {Janka}, \&
  {Heger}}]{MuJaHe12}
{M{\"u}ller}, B., {Janka}, H.-T., \& {Heger}, A. 2012{\natexlab{a}}, ApJ, 761,
  72

\bibitem[{{M{\"u}ller} {et~al.}(2012{\natexlab{b}}){M{\"u}ller}, {Janka}, \&
  {Marek}}]{MuJaMa12}
{M{\"u}ller}, B., {Janka}, H.-T., \& {Marek}, A. 2012{\natexlab{b}}, ApJ, 756,
  84

\bibitem[{{M{\"u}ller} {et~al.}(2013){M{\"u}ller}, {Janka}, \&
  {Marek}}]{MuJaMa12b}
{M{\"u}ller}, B., {Janka}, H.-T., \& {Marek}, A. 2013, ApJ, 766, 43

\bibitem[{M{\"u}ller \& Steinmetz(1995)}]{MuSt95}
M{\"u}ller, E. \& Steinmetz, M. 1995, Comp. Phys. Comm., 89, 45

\bibitem[{{Murphy} \& {Burrows}(2008)}]{MuBu08}
{Murphy}, J.~W. \& {Burrows}, A. 2008, ApJ, 688, 1159

\bibitem[{{O'Connor} \& {Ott}(2011)}]{OcOt11}
{O'Connor}, E. \& {Ott}, C.~D. 2011, \apj, 730, 70

\bibitem[{Rampp \& Janka(2000)}]{RaJa00}
Rampp, M. \& Janka, H.-T. 2000, ApJ, 539, L33

\bibitem[{{Reddy} {et~al.}(1998){Reddy}, {Prakash}, \& {Lattimer}}]{RePrLa98}
{Reddy}, S., {Prakash}, M., \& {Lattimer}, J.~M. 1998, Phys. Rev. D, 58, 013009

\bibitem[{{Scheck} {et~al.}(2006){Scheck}, {Kifonidis}, {Janka}, \&
  {M{\"u}ller}}]{ScKiJa06}
{Scheck}, L., {Kifonidis}, K., {Janka}, H.-T., \& {M{\"u}ller}, E. 2006, A\&A,
  457, 963

\bibitem[{{Smartt}(2009)}]{Smar09}
{Smartt}, S.~J. 2009, ARA\&A, 47, 63

\bibitem[{{Suwa} {et~al.}(2010){Suwa}, {Takiwaki}, {Whitehouse},
  {Liebend{\"o}rfer}, \& {Sato}}]{SuKoTa10}
{Suwa}, Y., {Takiwaki}, T., {Whitehouse}, S.~C., {Liebend{\"o}rfer}, M., \&
  {Sato}, K. 2010, PASJ, 62, L49

\bibitem[{{Takiwaki} {et~al.}(2012){Takiwaki}, {Kotake}, \& {Suwa}}]{TaKoSu12}
{Takiwaki}, T., {Kotake}, K., \& {Suwa}, Y. 2012, ApJ, 749, 98

\bibitem[{{Thompson} {et~al.}(2005){Thompson}, {Quataert}, \&
  {Burrows}}]{ThQuBu05}
{Thompson}, T.~A., {Quataert}, E., \& {Burrows}, A. 2005, ApJ, 620, 861

\bibitem[{{Woosley} \& {Heger}(2007)}]{WoHe07}
{Woosley}, S.~E. \& {Heger}, A. 2007, Phys. Rep., 442, 269

\bibitem[{{Woosley} {et~al.}(1994){Woosley}, {Wilson}, {Mathews}, {Hoffman}, \&
  {Meyer}}]{WoWiMa94}
{Woosley}, S.~E., {Wilson}, J.~R., {Mathews}, G.~J., {Hoffman}, R.~D., \&
  {Meyer}, B.~S. 1994, ApJ, 433, 229

\end{thebibliography}

\end{document}